\documentclass[]{elsarticle}
\usepackage{lineno}
\usepackage{amsmath}
\usepackage{gensymb}
\usepackage{braket}
\newcommand{\um}{\,\mu\text{m}}
\usepackage{xcolor}
\usepackage{ulem}
\title{Reflectionless propagation of beams through a stratified medium}
\author[1]{Sounak Sinha Biswas}
\author[2]{Ghanasyam Remesh}
\author[2]{Venu Gopal Achanta}
\author[1,3]{Ayan Banerjee}
\author[1,3]{Nirmalya Ghosh}
\author[1,4,5]{Subhasish Dutta Gupta\corref{cor1}}
\affiliation[1]{organization={Department of Physical Sciences, Indian Institute of Science Education and Research (IISER) Kolkata},
	city={Mohanpur}, postcode={741246}, country={India}}
\affiliation[2]{organization={Department of Condensed Matter Physics and Material Science, Tata Institute of Fundamental Research}, city={Mumbai}, postcode={400005}, country={India}}
\affiliation[3]{organization={Center of Excellence in Space Sciences India, Indian Institute of Science Education and Research (IISER) Kolkata},	city={Mohanpur}, postcode={741246}, country={India}}
\affiliation[4]{organization={Tata Centre for Interdisciplinary Sciences, TIFRH}, city={Hyderabad}, postcode={500107}, country={India}}
\affiliation[5]{organization={Indian Institute of Technology (IIT) Jodhpur}, city={Jodhpur},postcode={342030} , country={India}}
\cortext[cor1]{Corresponding author: sdghyderabad@gmail.com}

\begin{document}
\begin{abstract}
Reflectionless potentials following the prescription of Kay and Moses allow for total transmission of incoming waves of any kinetic energy. The optical analogue of such potentials occur as dielectric stratified media that can offer null reflectivity and near total transmission over a large range of incidence angles and wavelengths. In a previous work (S. Dutta Gupta and G. S. Agarwal, \textit{Opt. Express} 15, 9614-9624, 2007), this was demonstrated for linearly polarized plane waves. We extend the earlier work valid for plane waves to structured beams to show near-total transmission of beams across the reflectionless dielectric profile. The analysis is based on the angular spectrum decomposition treating the  beam  as a collection of plane waves.  Gaussian and Laguerre-Gaussian beams are shown to be transmitted through the film with $ <1\% $ reflection in most scenarios. We also discuss the superlative performance of our proposed profile in preserving the beam shape during transmission comparing these results to a conventional $ \lambda/2 $ antireflection coating.
\end{abstract}
\maketitle

\section{Introduction}
Reflection occurs whenever an electromagnetic field encounters an index inhomogeneity. In many practical applications such as microscopes or
telescopes, reflection needs to be suppressed in order to increase
the optical throughput. Back reflections from various optical
components such as lenses or waveplates are often undesirable
in experimental measurements. Suppressing unwanted reflections
can be also be important since stray beams can be a safety
hazard when working with high powered lasers. This issue is often mitigated by using antireflection coatings, the simplest kinds of which can be made with a film thickness of integral multiples of $ \lambda/2 $ or $ \lambda/4 $, depending on the refractive indices of bounding media. Such designs exploit the destructive interference between consecutive reflected rays from film and substrate interfaces. However, they are applicable only for a specific wavelength, incidence angle and polarization. There have been several studies on how to minimize reflection depending on whether the systems under study are lossless or lossy. As far as the former is concerned, there have been recent investigations leading to suppression of reflection
using techniques such as   variable-index thin films \cite{Sankur1984,Poitras2004,Tang1997} and sub-wavelength structured surfaces \cite{Kintaka2001, Lalanne1997,Li2018} (also see \cite{Baryshnikova2015} and references therein). Some newer techniques have also emerged which use metamaterials to realize universal impedance matching \cite{Im2018}. 
\par
Since the pioneering work of Kay and Moses \cite{Kay1956} on reflectionless potentials, there have been interesting demonstrations of total transmission of wave packets through their two-parameter realization \cite{Kiriushcheva1998, Lekner2007}, otherwise known as the Poschl-Teller ($sech^2$) potential. Such soliton potentials have been studied in detail in model computations and their experimental realizations \cite{Sukhorukov2010, Szameit2011}. The quantum mechanical interpretation ensures the total transparency of the potential with respect to an incoming particle irrespective of the initial kinetic energy. The optical analogue was studied in detail both theoretically and experimentally keeping in view the practical applications  \cite{Gupta2007,Thekkekara2014} as antireflection coatings.  It was shown that a stratified medium with reflectionless potential profile can ensure omnidirectional and broadband near-complete transmission of plane waves for both TE and TM polarizations. In the context of short pulse propagation through such a profile intriguing effects of negative group velocity dispersion and its role on supercontinuum generation were also explored \cite{Lekner2007, Vaity2021}. 
\par
The omnidirectional character of the reflectionless profile motivated us to look at the scattering (reflection and transmission) of a finite beam (retaining its vector character) through the inhomogeneous film representing the profile with or without substrate. It must be stressed that most of the previous studies on RP's do not  consider passage of structured light. Recall that by virtue of the angular spectrum decomposition a beam can be considered as a collection of plane waves with wave vectors on and around the central wave vector direction. Recently such an approach was applied to show giant Goos-H\"anchen shift in a stratified medium supporting bound states in continuum \cite{Biswas2023}. In this paper, Gaussian and Laguerre-Gaussian beams were shown to be transmitted through the film with $ <1\% $ reflection from the film in most scenarios. We also discussed the superlative performance of our proposed profile in preserving the beam shape during transmission, and compared these results to a conventional $ \lambda/2 $ antireflection coating. Our study focuses only on the Hermitian systems where our target is to enhance the optical throughput  through broadband omnidirectional transparency. In this context the KM profile is unique since it is the only justifiable model with full mathematical rigor. However there have been several studies on non-Hermitian systems just to suppress reflection exploiting Kramers-Kr\"onig media \cite{Horsley2015},or Mie resonances \cite{Spinelli2012} and other systems and phenomena \cite{Ye2017,Makris2020, Tzortzakakis2020}, some of which may have application potentials.
\par
Recently, structured light has come to occupy a prominent position in optical research \cite{Andrews2011,Angelsky2020,RubinszteinDunlop2016,Saha2019}. Fields with inhomogeneities in amplitude, phase, polarization and other parameters are being studied in detail, revealing a multitude of results interesting for both fundamental physics \cite{Bliokh2015,Bliokh2015a,Allen1992a,Remesh2022,Kumar2022} and practical applications like metamaterials \cite{Xu2016,Litchinitser2012}, trapping \cite{Woerdemann2012}, where manifestations of the spin-orbit interaction of light are observed as well \cite{Roy2022,Kumar2024}, imaging \cite{Geng2011,Gustafsson2000} and quantum information processing \cite{Berkhout2010,Langford2004}. Finding closed-form expressions of scattered fields for arbitrary structured beams is not feasible, therefore we developed a generic strategy based on the work of Bliokh and Aiello \cite{Bliokh2013} with some modifications to handle normal incidence and beams with larger transverse $ k $-spread. The method was applied  to map the intensity profiles of Gaussian beams reflected from a symmetric multilayered medium supporting bound states in continuum \cite{Biswas2023}. Examination of the spatial intensity profiles of beams can also help us to visually assess if they undergo any distortion. For instance, Laguerre-Gaussian beams can get severely deformed on passing through dielectric interfaces \cite{Bliokh2013,Okuda2008}, which can be avoided if Kay-Moses RPs are used. Finally, we compare our results with a traditional destructive interference-based antireflection coating. To the best of our knowledge, such a study involving structured light through reflectionless potentials, has not been attempted before, our main finding leading to the conclusion that the RP's are almost as good for structured beams as they are for plane waves.
\par
The organization of the paper is as follows. We discuss a general formulation of the problem in section \ref{sec:formulation}, with a brief description of the refractive index profile designed using the Kay-Moses prescription. We outline the angular spectrum method to calculate scattered beam shapes. In section \ref{sec:results}, we look at intensity profiles of reflected and transmitted Gaussian beams to assess the performance of RPs. We also look at the evolution of Laguerre-Gaussian beams in terms of its shape and intensity to compare RPs with traditional $ \lambda/2 $ antireflection coatings. We end with a summary of main results and future outlook in Conclusions.
\section{Formulation of the Problem}\label{sec:formulation}
In this section, we begin by laying out the formalism of Kay and Moses \cite{Kay1956} for calculating reflectionless potentials and mapping them to optical systems. This has been discussed in detail in a previous work \cite{Gupta2007}, hence we briefly go over the important points. The analogy between the optical and quantum mechanical systems in the context of one dimensional scattering is now well understood \cite{Kay1956,Gupta2016}. For example, for a TE-polarized plane monochromatic wave incident on a stratified medium with permittivity varying as $ \epsilon(z) $ with $ xz $ plane being the plane of incidence, the propagation  equation can be written as 
\begin{equation}\label{eq1}
	\frac{d^2\Psi}{dz^2} + (k_0^2 \epsilon(z) - k_x^2)\Psi=0,
\end{equation}
where $ k_0 = \omega/c $ is the vacuum wave vector, $ k_x=k_0\sqrt{\epsilon_s}\sin\theta $, and $\theta$ is the angle of incidence. $\epsilon_s$ is the background permittivity. The above Helmholtz equation can be mapped on to the stationary Schr\"odinger equation by  
introducing the potential $ V(z) $ and the energy $ E $ as follows
\begin{eqnarray}\label{eq23}
	V(z) &= &k_0^2\epsilon_s-k_0^2\epsilon(z),\\
	E &=& k_0^2 \epsilon_s \cos^2\theta,
\end{eqnarray}
The potential $V(z)$ is said to be reflectionless if a wave with arbitrary energy $ E $ completely passes through the potential. 
Kay and Moses provided a general prescription for finding such RPs in the context of inverse scattering theory \cite{Kay1956}. Assuming positive arbitrary constants $ A_1, A_2, \dots, A_N $ and $ \kappa_1, \kappa_2, \dots, \kappa_N $, we consider the following set of simultaneous linear equations
\begin{equation}\label{eq4}
	\sum_{j=1}^{N}M_{ij} f_j(z) = -A_i e^{\kappa_i z},\, \quad M_{ij} = \delta_{ij} + \frac{A_i e^{(\kappa_i + \kappa_j)z} }{\kappa_i + \kappa_j}.
\end{equation}
The potential turns out to be \cite{Kay1956}
\begin{equation}\label{eq5}
	V(z) = -2\frac{d^2}{dz^2}[\log(D)],
\end{equation}
where $ D = \det(M_{ij}) $.
Eq(2) then  leads to the  refractive index profile for the optical equivalent $ n(z) =\sqrt{\epsilon(z)}$ 
as follows
\begin{equation}\label{eq6}
	n^2(z) = n_s^2 + \frac{2}{k_0^2}\frac{d^2}{dz^2}[\log(D)].
\end{equation}
Retention of only one pair of parameters $ A_1 $ and $ \kappa_1 $, with $ A_1=2\kappa_1 $ yields the well known Poschl-Teller potential with $sech^2$ profile \cite{Kay1956, Kiriushcheva1998}. By adjusting the parameters $ A_i $ and $ \kappa_i $, we can calculate increasingly complicated refractive index profiles as per our requirements. Note that in a realistic case, the RP described in Eq. \ref{eq6} has to be deposited on a substrate. To incorporate this effect, we build the profile on a smooth $ \tanh $ ramp as follows
\begin{equation}\label{eq7}
	n^2(z) = n_{s_1}^2 + \frac{2}{k_0^2}\frac{d^2}{dz^2}[\log(D)] + \frac{n_{s_2}^2 - n_{s_1}^2}{2}(1+\tanh(\kappa_1 z)),
\end{equation}
where $ n_{s_1} $ and $ n_{s_2} $ are the refractive indices of the media on the left and right sides of the reflectionless profile.
\par
It is important to note that all steps in the above calculation have been carried out for TE polarization and a similar feat is not possible to achieve for TM waves since they cannot be recast in the form of Schr\"odinger equation. However, the profile constructed for TE polarization performs sufficiently well for TM-waves, as reported in \cite{Gupta2007}. Thus, the profile in Eq. \ref{eq6} acts as an acceptable form for a  RP for  arbitrarily polarized plane waves. Of course, in a real situation the profile must be truncated resulting in a  a finite width, but even so it performs well for large angular domains for a broad spectral range \cite{Gupta2007}. As will be shown below, the omnidirectional broadband features of RP makes it suitable for near reflectionless transmission for structured beams as well.
\par
Unlike a plane wave having a distinct wave vector, a beam has a spread of wave vectors about the central wave vector direction. The narrower is the beam in the transverse spatial plane, broader is the spread in the $k-$domain. We now describe a generic formalism developed to simulate the spatial intensity profile of a beam incident on a stratified medium. This approach, also used in a previous work \cite{Biswas2023}, is based on the technique developed by Bliokh and Aiello \cite{Bliokh2013} for a single interface, with two major improvements. Firstly, we do away with a strict paraxial approximation and perform a more accurate angular spectrum decomposition, rendering the theory applicable to normal incidence for beams with larger transverse spread in the momentum domain. Secondly, we calculate exact reflection and transmission coefficients for the stratified medium using the transfer matrix method, as opposed to using a first-order Taylor expansion. We now outline the important steps to produce the beam shapes of reflected and transmitted beams. 
\par
We start, for example, with the normalized spectrum of a Gaussian beam given by \cite{Bliokh2013,Biswas2023}
\begin{equation}\label{eq8}
	\ket{\mathbf{E}_i} = \frac{w_0}{\sqrt{2\pi}}\exp(-(k_x^2+k_y^2)w_0^2/4) (A_p\hat{e}_p + A_s\hat{e}_s), 
\end{equation}
where $ w_0 $ is the beam waist and the second parenthesis describes the polarization. Assume that the central wave vector of the beam is incident at an angle $ \vartheta_i $ (with $ \vartheta_r $ and $ \vartheta_t $  being the reflection and transmission angles respectively). Angular decomposition of the off-axis wave vectors can be performed by using the following expressions from \cite{Zhu2021} 
\begin{equation}\label{eq9}
	\theta_i = \tan^{-1} \left( \frac{\sqrt{k_y^2 + (k_x\cos \vartheta_i + k_z\sin\vartheta_i)^2}}{-k_x\sin\vartheta_i + k_z\cos\vartheta_i} \right),
\end{equation}
\begin{equation}\label{eq10}
	\phi_i = \tan^{-1}\frac{k_y}{k_x \cos\vartheta_i+k_z\sin\vartheta_i},
\end{equation}
where $k_z=\sqrt{k_0^2-(k_x^2+k_y^2)}$. These polar and azimuthal angles are also calculated for scattered beams to carry out the rotational transformation from the beam frame to that of $ p $ and $ s $ modes. The computation of the spectral profiles of the scattered beams requires reflection and transmission coefficients for  both $p$ and $s$ polarizations. These coefficients ($a_p$ and $a_s$) are determined for the RP film using the transfer matrix method \cite{Gupta2016} for each of these individual harmonics. Finally the output spectra can be calculated as follows \cite{Bliokh2013}
\begin{equation}\label{eq11}
	\ket{\mathbf{E}_a} = U^\dag_a F_a U_i  \ket{\mathbf{E}_i},
\end{equation}
where
\begin{eqnarray}
	U_a = \hat{R}_y(\theta_a) \hat{R}_z(\phi_a) \hat{R}_y(-\vartheta_a),\label{eq12}\\
	F_a = \mbox{diag}(a_p, a_s),\quad a=r,t,\label{eq13}
\end{eqnarray}
with subscripts $i, r,t$, indicating the relevant quantities for incident, reflected and transmitted beams respectively. $ \hat{R}_{x,y,z} $ are the $ x,y,z $ rotation matrices in the lab frame. Inverse Fourier transforms of the spectra given by Eq. \ref{eq11} yields the respective beam shapes. In the next section, we discuss the effectiveness of reflectionless profiles for near perfect transmission of Gaussian and Laguerre-Gaussian beams. For the latter, we replace the expression in Eq. \ref{eq8} with the following \cite{Bliokh2013}
\begin{eqnarray}
	&\ket{\mathbf{E}_i} \propto \frac{w_0}{\sqrt{2\pi}}\exp(-(k_0 w_0)^2 \theta_z^2/4) \theta_z^{|l|} e^{il\phi+ik_0(1-\theta_z^2/2)z}\\ \nonumber
	&\times (A_p\hat{e}_p + A_s\hat{e}_s),\label{eq14}
\end{eqnarray}
where $\theta_z = \sqrt{k_x^2+k_y^2}/k_0$, $\phi =\tan^{-1}(k_y/k_x)$ and the azimuthal phase factor $ e^{il\phi} $ represents an optical vortex of charge $ l = 0,\pm 1, \pm 2, \dots $.
\section{Results and Discussion}\label{sec:results}
We start by first specifying the parameters of the reflectionless potentials used for the numerical calculations. As mentioned earlier for only one pair of parameters $ A_1 $ and $ \kappa_1 $, with $ A_1=2\kappa_1 $, we have the standard Poschl-Teller profile. To achieve low reflectivity for large ranges of wavelength and angle of incidence, we employ a four-parameter family, similar to \cite{Gupta2007}, $ A_1=11,A_2=A_3=A_4=3 $ and $\kappa_1=5.5, \kappa_2=0.1, \kappa_3=1.0, \kappa_4=9.0 $. The potential described in Eq. \ref{eq6} or Eq. \ref{eq7} is infinite in extent and they were  truncated  to the domain  $ -3 \um <z<3\um $ retaining the essential features. The incident Gaussian beam spectrum is calculated using Eq. \ref{eq8}, with a wavelength $\lambda=1.06\um $  and beam waist $ w_0=10\lambda $. This design wavelength is suited for the parameter family we are using. We study the reflection characteristics of the RP film for both cases, with and without substrate. In the latter case, the RP is deposited on a substrate with refractive index $ n_{s2}=1.4 $. $n_s$ or $n_{s1}$  is taken to be 1 in all the calculations.
\par
\begin{figure}[tp]
	\centering
	\includegraphics[width=\linewidth]{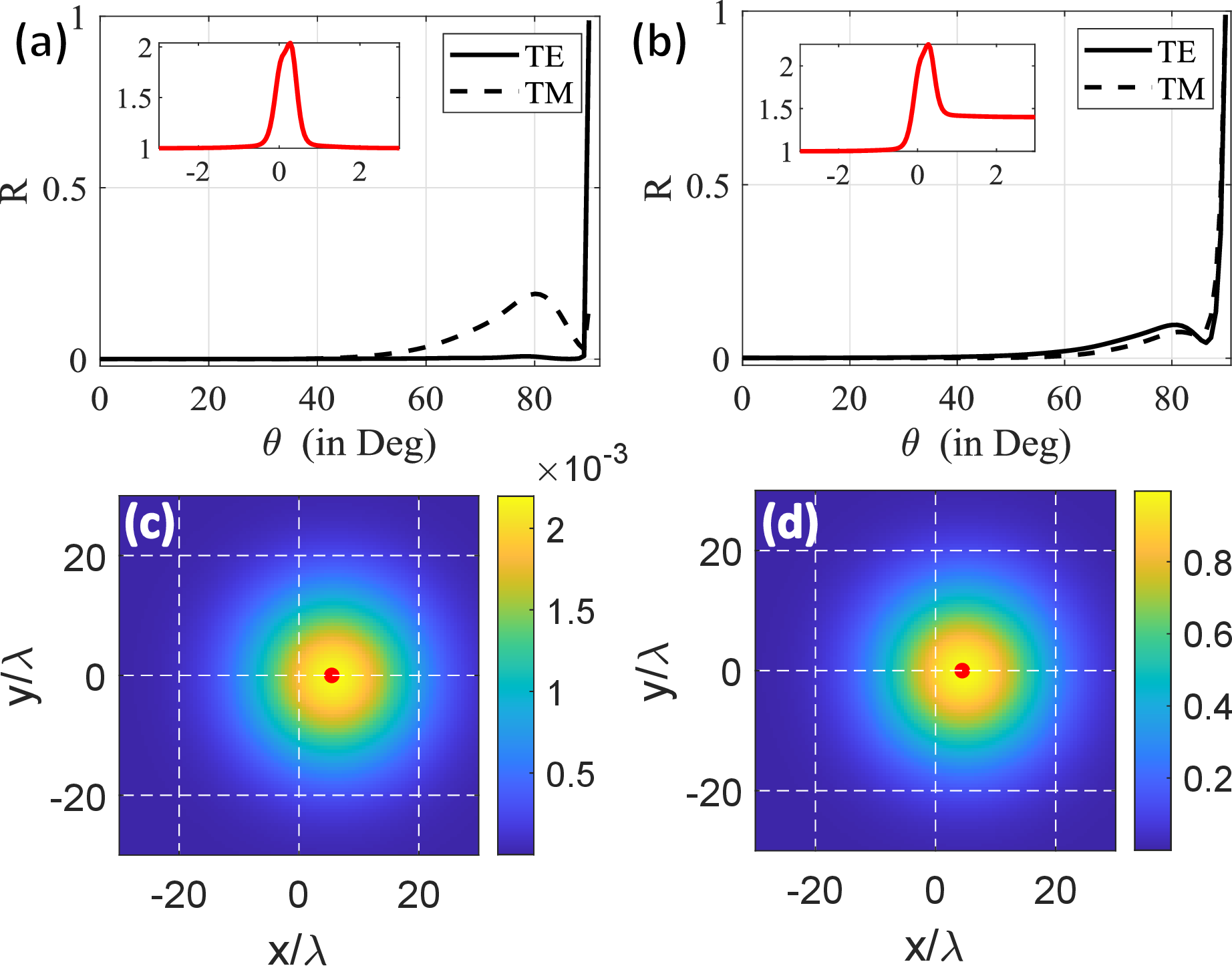}
	\caption{Intensity reflection coefficient R for Kay-Moses RP (a) without and (b) with  substrate, shown for TE and TM polarized plane waves at a wavelength of $1.06\um$. The refractive index profiles are plotted as inset in the corresponding figures. Spatial intensity profiles of (c) reflected  and (d) transmitted Gaussian beams of the same wavelength incident at an angle of $ 30\degree $ on the RP with substrate. The red dot indicates the beam centroid.}
	\label{fig:figure1}
\end{figure}

Our strategy is first to look at the reflection and transmission for plane polarized waves and identify the angle and wavelength ranges where the RP's are most effective. Though most of these results are discussed in an earlier paper (\cite{Gupta2007}), this serves as a proper reference to understand the results pertaining to the beams. The same structures are then probed for the reflection and transmission of the structured beams. We begin by examining the intensity reflection coefficients as a function of angle of incidence calculated for plane waves (as in \cite{Gupta2007}) of wavelength $1.06\um$. Fig. \ref{fig:figure1}(a) shows a predominantly flat near zero reflection coefficient for the RP film without any substrate, deviating only at larger, especially grazing,  angles for TE waves. The behaviour for TM waves is also similar, but has a noticeable peak at $ 80\degree $. This is to be expected since the formulation applies exactly only for TE polarization. However, for the RP on a substrate (see Fig. \ref{fig:figure1}(b)) the response of both TE and TM waves are nearly the same. The refractive index profiles in both the cases have been shown as insets in the corresponding figures, 
\par
Next, we consider a Gaussian beam with TE polarization incident on a RP with substrate at an angle of $ \theta=30\degree$ for a wavelength of $1.06\um$. The reflected and transmitted beam shapes calculated using Eq. \ref{eq11} are shown in Fig. \ref{fig:figure1}(c,d). Note that the intensity values here have been normalized with respect to the maximum intensity of incident beam. The reflected beam intensity illustrates the negligible reflectivity of the Kay-Moses RP for realistic Gaussian beams. The Gaussian beam is transmitted almost completely through this film, without undergoing any discernible distortion. A large  Goos-H\"anchen shift is observed for reflected and transmitted beams, the red dot locating the centroid of the beam.
\par
\begin{figure}[tp]
	\centering
	\includegraphics[width=\linewidth]{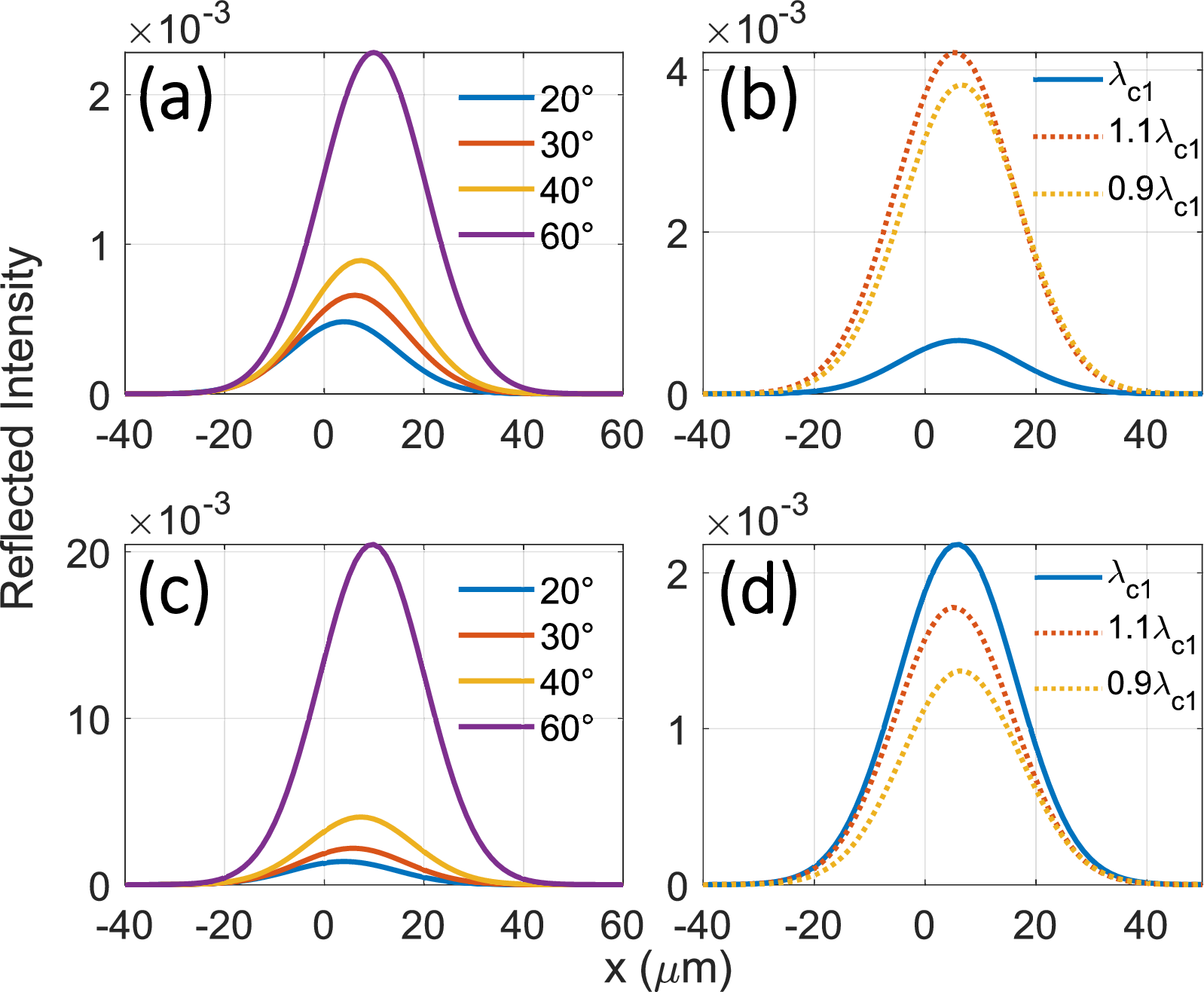}
	\caption{Spatial intensity profile cross section of reflected Gaussian beams from a RP film without substrate as a function of $ x $ for $ y=0 $, calculated for different (a) angles and  (b) wavelengths. (c) and (d) shows the same for a RP on a substrate. The design wavelength $\lambda_{c1}$ for the RP is $1.06\um$.}
	\label{fig:figure2}
\end{figure}

Note that although the film was designed for wavelength $ \lambda_{c1}=1.06\um $, its performance is broadband. This has been demonstrated in the previous study \cite{Gupta2007} in the regime of plane waves. We continue the analysis of this feature of RPs for beams. In Fig. \ref{fig:figure2}, we examine the performance of the RP at different conditions of wavelength, angle of incidence and presence of substrate. We calculate the reflected beam shape for each case and take its cross-section at $ y=0 $ to study the trend of intensity. For simplicity, we focus only on TE polarized Gaussian beams in this study. Fig. \ref{fig:figure2}a (Fig. \ref{fig:figure2}b) shows the performance of film without substrate for varying angle of incidence (wavelength) at a fixed wavelength of 1.06 $\um$ (angle of $\theta = 30\degree $) . We see that the reflected intensity normalized with respect to the incident intensity maxima is of the order of $ 10^{-3} $, for all angles and wavelengths considered. This is a noteworthy result, for it shows that the concept of reflectionless potentials developed for plane waves works just as flawlessly for finite beams. In Fig. \ref{fig:figure2}(c,d) we examine the same results for a more realistic case of film with substrate. For the different angles of incidence, there is approximately an order increase in the intensity of reflected light as compared to the film without substrate. Varying the frequency does not seem to effect the performance of the film as much, although we see a slight increase in the reflected light.
 \par
In order to emphasize the uniqueness of Kay-Moses RPs, we compare them to conventional antireflection films which use the concept of destructive interference between consecutive reflected rays to yield low reflectivity in the medium of incidence. The usual AR coatings use a $\lambda/4$ slab with refractive index being equal to the geometric mean of the refractive indices of the ambient media ( $n_2=\sqrt{n_1 n_3}$). Recall that the peak refractive index of our RP is higher than that of the substrate (see Fig. \ref{fig:figure1}(b) inset) and such a profile can not be compared with a stepwise increasing refractive index profile of the AR coating. Instead we compare the performance of the RP with a uniform high index  $\lambda/2$ film with $ n_2  > n_1,n_3 $ ensuring destructive interference in the medium of incidence . In such a case, the thickness of the film has to be integral multiples of $ \lambda/2n_2\cos\theta_2 $ where $\theta_2$ is the angle of refraction inside the film \cite{Born2013}. Clearly, this design is angle and wavelength specific unlike the Kay-Moses potential. Moreover, it needs to be stressed that its offered reflectivity does not reach zero, which is expected from destructive interference, unless $ n_1=n_3 $ \cite{Born2013}.  Hence, the two kinds of films can only be compared for the ideal case without a substrate. Since the films are to be grown on a substrate we present results for cases without and with substrate.  In order to avoid confusion we refer to such films as uniform film (UF). In all of our calculations the value of $n_2$ is chosen such that the area under the two profiles, namely, the UF and the RP match.
\begin{figure}[htp]
	\centering
	\includegraphics[width=\linewidth]{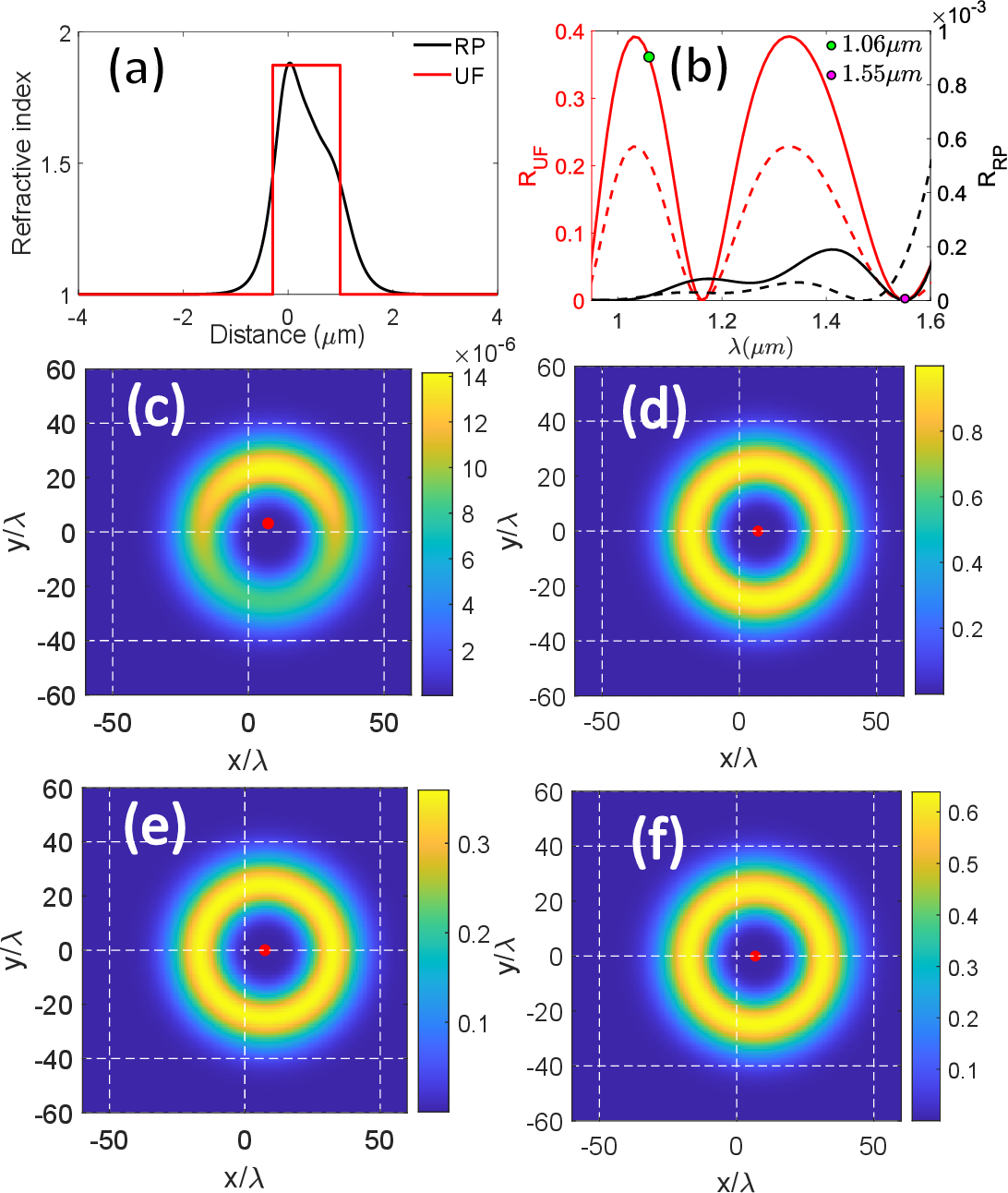}
	\caption{(a) Index profiles of RP and UF designed for $ 1.55\um $, without considering a substrate. (b) Reflection coefficients at $\theta=30\degree$ for the structures shown in (a), following the same colour scheme (Note that reflection coefficients for UF and RP are plotted on different axis with different scales). The solid and dashed lines denote TE and TM polarizations. Spatial intensity profiles of a (c) reflected (d) transmitted LG beam with $\lambda = 1.06\um$ incident on a substrateless RP film at $ 30\degree$. (e) and (f) shows the same for a UF profile. The red dot indicates the beam centroid. }
	\label{fig:figure3}
\end{figure}
\par
The RP is now constructed using a 3-parameter family with $ A_1=11,A_2=8,A_3=5.5 $ and $\kappa_1=5.5, \kappa_2=4, \kappa_3=2.25 $ at $ 1.55\um $. Such a parameter selection gives us almost similar low reflectivities for both polarizations. We begin by considering no substrate and calculate the thickness and index of the UF profile as mentioned previously. Both structures are designed at $\lambda_{c2}=1.55\um$, with the UF profile at $30\degree$. They are shown in Fig. \ref{fig:figure3}(a) with appropriate legends. To compare their anti-reflective performances, we will study the structures away from the design wavelength at $\lambda=1.06\um$ and $\theta=30\degree$. We plot plane-wave reflectivity as a function of wavelength in Fig. \ref{fig:figure3}(b). Right away, we observe that the RP has almost overlapping flat curves for both polarizations, throughout the wavelength range in consideration. On the other hand, the UF profile, despite offering zero reflectivity at $1.55\um$, reflects considerably at $1.06\um$. We continue this analysis using a TE-polarized Laguerre-Gaussian beam with vortex charge $ l=3 $ (see Eq. \ref{eq14}). For a Kay-Moses RP, the reflected beam profile of the LG beam at a wavelength of $ 1.06\um $ incident at an angle of $30 \degree$ has a  very low intensity.  The corresponding reflected and transmitted beam shapes are shown in Fig. \ref{fig:figure3}(c,d).
\par
Since we do not consider a substrate, the uniform profile shows zero reflectivity at $ 30\degree $ at the design wavelength. However, such designs are specific to angle and wavelength. We demonstrate the failure of a uniform profile to exhibit antireflection behaviour at other wavelengths, by using an LG beam at $ 1.06\um $, as shown in Fig. \ref{fig:figure3}(e,f), where the reflected beam profile can be seen to be considerably more intense.
\begin{figure}[htp]
	\centering
	\includegraphics[width=\linewidth]{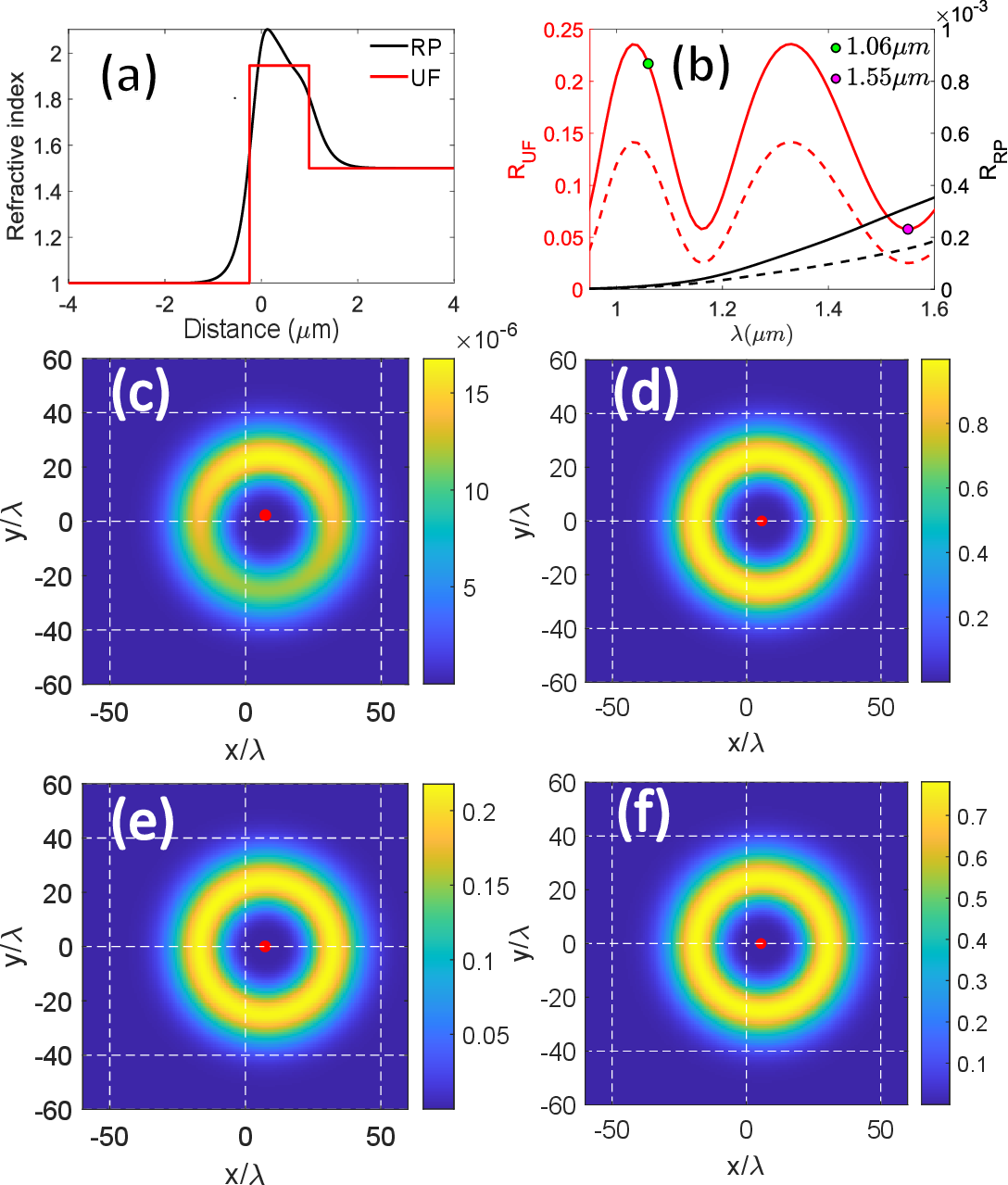}
	\caption{  (a) Index profiles of a RP and UF designed for $ 1.55\um $, on a substrate of $n_{s2}=1.5$. (b) Reflection coefficients at $\theta=30\degree$ for the structures shown in (a), following the same colour scheme (Note that reflection coefficients for UF and RP are plotted on different axis with different scales). The solid and dashed lines denote TE and TM polarizations. Spatial intensity profiles of a (c) reflected (d) transmitted LG beam with $\lambda = 1.06\um$ incident on a RP film on substrate at $ 30\degree$. (e) and (f) shows the same for a UF profile. The red dot indicates the beam centroid.}
	\label{fig:figure4}
\end{figure}
\par
Incorporating substrate effects shows us similar results. The RP and UF profiles are shown in Fig. \ref{fig:figure4}(a), both grown on a substrate of refractive index $ 1.5 $. We see similar results as before in the $R(\lambda)$ plot in Fig. \ref{fig:figure4}(b), with the additional observation that the reflectivity of UF profile does not reach zero in this case, even at the design angle. On the other hand, the RP has a near zero flat reflectivity profile for all wavelengths. As before, we examine reflected/transmitted profiles of TE-polarized LG beams at $ 1.06\um $, incident at $ 30\degree $.
In case of the RP (Fig. \ref{fig:figure4}(c,d)), we observe low reflectivity as expected, and for the UF profile, reflected intensity rises considerably at this wavelength(see Fig. \ref{fig:figure4}(e,f)). Thus, we have established the uniqueness and versatility of Kay-Moses RPs over conventional antireflection films, even when the film is grown on a substrate.
\par
We have extensively tested the performance of both profiles, and sum it up as follows. We see that the RP profile works consistently at all wavelengths below $ 1.55\um $, up to high angles of incidence ($ \approx 70\degree $) for both TE and TM polarizations. On the other hand, the UF profile does not offer low reflectivity away from the design angle for both polarizations, and fails completely when a different wavelength is used.
\par
\begin{figure}[htp]
	\centering
	\includegraphics[width=\linewidth]{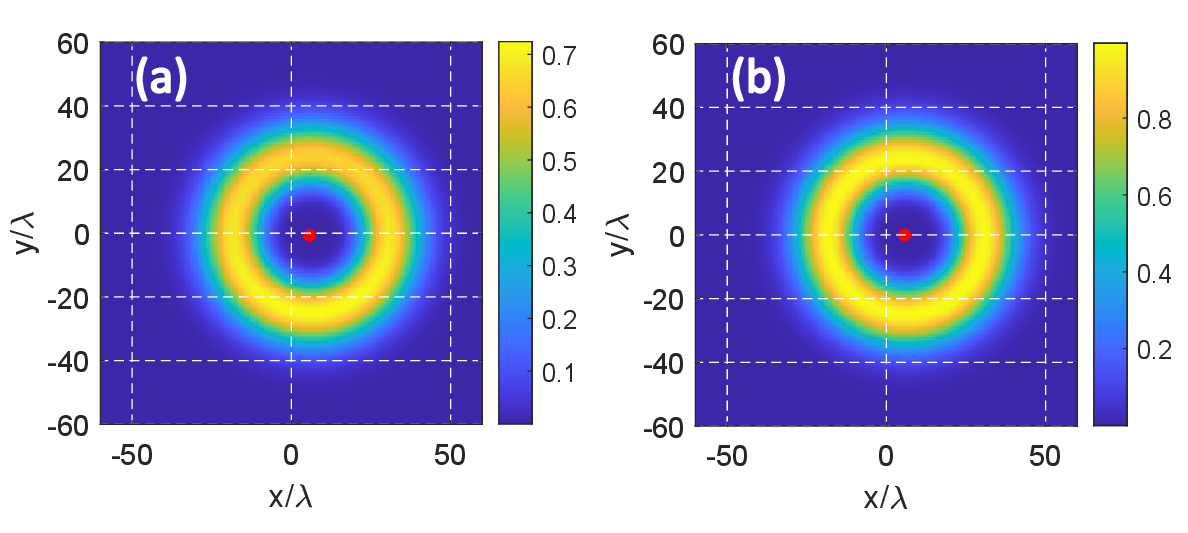}
	\caption{Spatial intensity profile of a TE-polarized LG beam at $1.55\um$ with incidence angle $60\degree$ transmitted through a uniform (a) and reflectionless potential profile (b), both grown on a substrate of $n_s=1.5$. The red dot indicates the beam centroid.}
	\label{fig:figure5}
\end{figure}

An additional advantage of using Kay-Moses reflectionless potentials is that they allow the entire beam to pass through without any distortion unlike, for example, dielectric interfaces, which often cause orbital angular momentum-carrying beams to get deformed \cite{Okuda2008, Bliokh2013}. We demonstrate a specific case where the UF profile distorts the transmitted beam while the RP does not. In Fig. \ref{fig:figure5}, a TE-polarized LG beam at $1.55\um$ is considered to be incident on the UF and RP profiles at $60\degree$. Owing to the angle-specific design of the former, the transmitted beam gets deformed, with an IF shift in the $-y$ direction. On the other hand, the beam transmitted through the  RP undergoes negligible distortion, retaining the vector character of the beam. Note that both profiles here were grown on a substrate of refractive index $1.5$.
\section*{Conclusions}
To conclude, we proposed an inhomogeneous film based on the approach of Kay and Moses to obtain an omnidirectional, broadband reflectionless potential that is applicable in the context of structured light. We used a generic strategy, employing the angular spectrum decomposition to study the interaction of an arbitrary beam with a planar stratified medium. It was then applied to a Gaussian beam incident on the reflectionless potential, deposited on a glass substrate, for which we found low reflectivity for both TE and TM polarizations over a wide range of incidence angle and wavelength. We proceeded to compare the Kay-Moses potential to a uniform antireflection coating using Laguerre-Gaussian beams and demonstrated the former's wider applicability, owing to consistently low reflectivity at different wavelengths and incidence angles, irrespective of the polarization. Finally, we also showed that Kay-Moses profiles allow vector beams to pass through the medium without any change to their spatial character, which is often not the case with dielectric interfaces. All of these findings were arrived at by examining the spatial intensity profiles of the scattered beams. This study demonstrates the applicability of reflectionless potentials in increasing the optical throughput for structured light, which can hopefully open up novel research possibilities for both fundamental physics and practical applications.

\bibliography{refs}

\end{document}